\documentclass[twoside,11pt]{article}

\usepackage{blindtext}

\usepackage[abbrvbib, preprint]{jmlr2e}

\usepackage{natbib} % has a nice set of citation styles and commands
    
\usepackage{mathtools} % amsmath with fixes and additions
\usepackage{booktabs} % commands to create good-looking tables
\usepackage{tikz} % nice language for creating drawings and diagrams
\usepackage{amsmath}
\usepackage{graphicx}
\usepackage{multirow}
\usepackage{dsfont}
\usepackage{mathrsfs}
\usepackage{array}
\usepackage{xcolor}
\usepackage{float}
\usepackage{enumitem}   

\newcommand \po[1]{Y_i^{(#1)}}

\newtheorem{specification}{Specification}

\ShortHeadings{}{}
\firstpageno{1}

\begin{document}

\title{Targeting relative risk heterogeneity with causal forests}

\author{\name Vik Shirvaikar\textsuperscript{1} \email vik.shirvaikar@spc.ox.ac.uk \\
        \name Andrea Stor{\aa}s\textsuperscript{2} \email btis@novonordisk.com \\
        \name Xi Lin\textsuperscript{1} \email xi.lin@stats.ox.ac.uk \\ 
        \name Chris Holmes\textsuperscript{1} \email chris.holmes@stats.ox.ac.uk \\[1ex]
        \addr 
            \textsuperscript{1}Department of Statistics, University of Oxford
            \textsuperscript{2}Novo Nordisk A/S}

\maketitle

\begin{abstract}
The identification of heterogeneous treatment effects (HTE) across subgroups is of significant interest in clinical trial analysis. Several state-of-the-art HTE estimation methods, including causal forests, apply recursive partitioning for non-parametric identification of relevant covariates and interactions. However, the partitioning criterion is typically based on differences in absolute risk. This can dilute statistical power by masking variation in the relative risk, which is often a more appropriate quantity of clinical interest. In this work, we propose and implement a methodology for modifying causal forests to target relative risk, using a novel node-splitting procedure based on exhaustive generalized linear model comparison. We present results from simulated data that suggest relative risk causal forests can capture otherwise undetected sources of heterogeneity. We implement our method on real-world trial data to explore HTEs for liraglutide in patients with type 2 diabetes.
\end{abstract}

\section{Introduction}

In clinical settings, we are often interested in exploring the evidence for heterogeneous treatment effects (HTE), or whether specific subgroups of the population respond differently to a treatment under investigation. This question has received significant attention in the causal inference literature over the past few decades, as it represents a key step in the direction of personalized medicine \citep{sechidis_watch_2024, lipkovich_modern_2024}.

More specifically, HTE discovery calls for nonparametric methods which can consider the dataset as a whole to identify subgroups of interest \citep{watson_graphing_2020}. Classical subgroup analysis, where potentially relevant covariates are pre-specified in a clinical trial protocol, may fail to detect strong but unexpected heterogeneity, and additionally raises concerns related to multiple testing \citep{cook_subgroup_2004}. For this reason, forest-based methods have become especially popular, building upon the seminal work of \citet{breiman_random_2001} to flexibly model high-dimensional interactions between covariates.

In particular, causal forests \citep{athey_recursive_2015, wager_estimation_2018, athey_generalized_2019} are a state-of-the-art approach for HTE estimation in real-world clinical trial analysis \citep{basu_characteristics_2018, athey_estimating_2019, raghavan_generalizability_2022}. While a standard decision tree recursively partitions the input data to maximize the variability in an outcome, a causal decision tree instead maximizes the variability in treatment effect by separating the treatment and control samples within each node. To adjust for confounding and directly estimate HTEs, causal forests orthogonalize the outcome and treatment propensity with respect to the covariates, leveraging ideas from the double/debiased machine learning literature \citep{chernozhukov_doubledebiased_2018}. 

However, the node-splitting criterion in causal forests targets heterogeneity in the additive or absolute Risk Difference (RD) between subgroups. In certain contexts, a preferable approach is to target heterogeneity in the multiplicative or relative Risk Ratio (RR). The RR has been found to extrapolate more effectively across populations \citep{deeks_issues_2002, furukawa_can_2002}, meaning that heterogeneity in the RR may be more indicative of a true difference in treatment efficacy \citep{sun_how_2014, spiegelman_evaluating_2017}.

In addition, the RD over-emphasizes individuals with a high baseline risk level \citep{kent_assessing_2010}. As a result, HTE estimates based on RD could be biased towards high-risk individuals, and spuriously identify prognostic covariates that are related to baseline risk, rather than predictive covariates that indicate treatment heterogeneity. Targeting the RR instead can improve statistical power, especially in scenarios where there is large variation in individual baseline risk, by weighting all individuals equally regardless of risk level.

We therefore aim to adjust the structure of causal forests such that the RR can be chosen as the quantity of interest. We present a novel method that uses exhaustive generalized linear model (GLM) comparison as the basis for the forest splitting rule. By fitting a GLM with an interaction term between the treatment and every possible candidate split in succession, we identify the split that induces the most significant heterogeneity in the treatment effect. Changing the link function of the GLM then allows for quantities other than the RD, such as the RR, to be targeted. Our approach is implemented as an update to the \texttt{grf} software package in \texttt{R} \citep{tibshirani_grf_2023}, available at \texttt{https://github.com/vshirvaikar/rrcf}. 

Previous partitioning approaches such as RECPAM \citep{ciampi_recpam_1988} have implemented an exhaustive model-based search, but never in the causal setting, or with a focus on RR. We additionally note the recent body of work on ``model-based forests'', which identify the optimal split by fitting a model within each node of a decision tree \citep{zeileis_model-based_2008, seibold_model-based_2016}. However, the mechanism and motivation of this approach are fundamentally different: model-based forests fit a single model within each node in order to simultaneously estimate prognostic and predictive effects, while we fit several models within each node for the purpose of identifying the optimal split. 

The remainder of this article is organized as follows. We present the problem setting in Section 2, and motivate the importance of relative risk in Section 3. In Section 4, we review existing forest-based methods for HTE estimation, and in Section 5, we present our proposed methodology and demonstrate how it fits into the causal forest framework. Section 6 validates our approach on simulated data. Section 7 provides an application to real-world data from the LEADER clinical trial, and Section 8 concludes the paper.

\section{Problem setting}

Consider a clinical trial with $n$ subjects, each with covariates $X_i$ and a binary treatment assignment indicator $W_i \in \{0,1\}$ for $i = 1, \ldots, n$. In the current work, we focus on the binary outcome setting with $Y_i \in \{0,1\}$. To estimate the causal effect of the treatment on the outcome, we adopt the potential outcomes framework of \citet{rubin_causal_2005}. Under this framework, each subject has potential outcomes $\po{1}$, representing the outcome if the subject were treated ($W_i = 1$), and $\po{0}$, representing the outcome if the subject were untreated ($W_i = 0$). The fundamental challenge in causal inference is that for each subject, we can only observe one of these outcomes depending on treatment assignment
$$Y_i = W_i \po{1} + (1 - W_i) \po{0}.$$

To estimate causal effects from observed data, we must therefore introduce certain key assumptions. The Stable Unit Treatment Value Assumption (SUTVA) states that there is no hidden variation in the treatment $W$, and no interference between the treatment outcomes for different subjects. Ultimately, this implies that each subject's outcome depends only on their own treatment status $\left[Y_i = \po{W_i}\right]$. We additionally assume ignorability or unconfoundedness, meaning that treatment assignment is independent of potential outcomes given the observed covariates $\left[(\po{1}, \po{0}) \perp\!\!\!\perp W_i \mid X_i\right]$. Finally, we assume positivity or overlap, where each subject has a nonzero probability of receiving either treatment $\left[0 < \mathbb{P}(W_i = 1 \mid X_i) < 1 \;\; \forall X_i\right]$.

Together, these assumptions allow us to identify various causal estimands for our treatment effect. Note that in the binary setting, expected outcomes can be interpreted as probabilities $\mathbb{E}\left[\po{W_i}\right] = \mathbb{P}\left[\po{W_i}=1\right].$ Two common outcome measures are the \emph{Risk Difference} (RD), which measures the absolute change in event probability due to treatment,
$$\tau_{RD} = \mathbb{E}\left[\po{1}-\po{0}\right] = \mathbb{P}\left[\po{1} = 1\right] - \mathbb{P}\left[\po{0} = 1\right],$$
and the \emph{Risk Ratio} (RR), which measures the relative change in event probability between the treated and untreated groups,
$$\tau_{RR} = \frac{\mathbb{E}\left[\po{1}\right]}{\mathbb{E}\left[\po{0}\right]} = \frac{\mathbb{P}\left[\po{1} = 1\right]}{\mathbb{P}\left[\po{0} = 1\right]}.$$
Though we will not focus on them here, other common measures used in the binary setting include the Odds Ratio (OR), which compares the odds of the event under treatment versus control and frequently appears in case-control studies; and the Number Needed to Treat (NNT), which contextualizes the RD using patient counts and is widely used in medical practice \citep{cook_number_1995, altman_practical_1999}.

The RD is also referred to as the average treatment effect (ATE) across the population. However, this does not capture variations in treatment response across different subgroups of the study population, motivating the need for heterogeneous treatment effect (HTE) estimation \citep{rothman_epidemiology_2012}. A typical target quantity for HTE estimation is then the expected RD for a subject with a given set of covariates
$$\tau_{RD}(x) = E[\po{1} - \po{0}|X_i = x]$$
or, equivalently, the conditional average treatment effect (CATE).

\section{The importance of relative risk}

In many practical applications, there are compelling reasons to prefer HTE estimation on a relative rather than an absolute scale, or to evaluate both in parallel. Instead of the CATE, the relevant estimand becomes the expected RR given covariates
$$\tau_{RR}(x) = \frac{E[\po{1}|X_i = x]}{E[\po{0}|X_i = x]},$$
which has been referred to as the conditional Risk Ratio (CRR) \citep{wang_measuring_2016}.

Relative effect measures such as the RR are often preferred due to their perceived stability across settings. Multiple meta-analyses of clinical trials have found that RR reductions extrapolate most reliably across different time periods and populations \citep{schmid_empirical_1998, furukawa_can_2002, deeks_issues_2002}. As a result, it has been argued that heterogeneity in the RR is more indicative of true underlying effect modification than heterogeneity in the RD or OR \citep{sun_how_2014, spiegelman_evaluating_2017}. We note that certain contrasting empirical settings have been highlighted, in which RD or OR estimates are likely more stable \citep{poole_is_2015, doi_controversy_2022}.

The RR also possesses several desirable theoretical properties. Under a range of plausible causal models, \citet{huitfeldt_shall_2022} and \citet{colnet_risk_2023} show that the RR remains stable between patient groups, provided that the standard RR is used if the treatment is beneficial, and the Survival Ratio (the RR reversed to count null events) is used if the treatment is harmful. This formulation is equivalent to the Switch Relative Risk of \citet{van_der_laan_estimation_2007}, and addresses long-standing concerns about the asymmetry of the RR first discussed in \citet{sheps_shall_1958}. \citet{stensrud_identification_2023} find that the RR is stable when exposure is partially unobserved, while \citet{piccininni_immune-selection_2025} find that the RR is stable when immune individuals are excluded from a study.

The RR presents further advantages in terms of explanation and interpretability. Relative effects often better align with how patients and practitioners perceive risk \citep{simon_understanding_2001, schechtman_odds_2002}. For example, expressions such as “five times higher risk” are frequently more intuitively meaningful than statements like “a 4\% increase in risk,” particularly when the outcome of interest is rare. In applied settings, it is generally recommended that relative and absolute effect measures are reported in parallel, in order to provide appropriate context and support accurate interpretation \citep{noordzij_relative_2017, colnet_risk_2023}.

Finally, RR-based estimands can be more appropriate in settings with substantial variation in baseline risk. For instance, \citet{watson_graphing_2020} present a malaria case study from the AQUAMAT clinical trial where patients' predicted baseline mortality risk varied from less than 1\% to greater than 80\%. In such cases, estimates based on differences in RD are often dominated by covariates associated with overall mortality. Individuals with a high pre-treatment risk may be ``overrepresented" in the final results, since a risk reduction from 40\% to 10\% (for example) is considered ten times as important as a reduction from 4\% to 1\%. While this weighting is appropriate when estimating a population-average effect, it may dilute statistical power for HTE detection by discounting individuals with a low pre-treatment risk level. Estimates based on RR would weight these reductions equally, leading to a more balanced detection of subgroup differences across the covariate distribution. 

Together, these motivations help to explain the widespread use of the RR in clinical research. A survey of 100 clinical trials published in the \emph{New England Journal of Medicine} between 2018 and 2020 found that the RR was employed approximately twice as often as the RD to report subgroup effects \citep{andersen_absolute_2021}. Despite this, state-of-the-art HTE estimation methods, such as causal forests, are still primarily designed to target heterogeneity on an absolute scale. Given the prominent role of the RR in medical decision-making, we aim to develop an HTE estimation method that specifically operates on a relative scale. In the following sections, we review existing forest-based approaches to HTE estimation, then describe our proposed modifications for targeting relative treatment effects.

\section{Review of existing approaches}

Tree-based methods non-parametrically partition the covariate space to adaptively model complex interactions, making them particularly well-suited for detecting HTEs, especially in high-dimensional settings. Classical decision trees and random forests \citep{breiman_classification_1984, breiman_random_2001} are widely used in supervised learning, where the partitioning process aims to optimize a variance-based loss function (for regression) or a Gini impurity criterion (for classification) with respect to the outcome. 

However, HTE estimation requires a different objective: rather than identifying covariates which explain variability in the outcome itself, the goal is to identify covariates which explain variability in the treatment effect, or in other words, \emph{differences in the differences} between the treated and untreated groups. Various modifications to recursive partitioning have been developed to achieve this; we discuss two key frameworks below.

\subsection{Causal forests}

Causal forests \citep{athey_recursive_2015, wager_estimation_2018} take a direct approach, specifying the ATE as the quantity of interest within a random forest framework. Suppose that we are evaluating whether to split on variable $Z$ at value $z$ at a given node in a decision tree. For some target quantity $\xi$, we would calculate the reduction in variance from the parent node to the resulting children, choosing the split that maximizes
$$\Delta V = \text{Var}(\xi_i) - p_L \cdot \text{Var}(\xi_i \mid Z_i < z) - p_R \cdot \text{Var}(\xi_i \mid Z_i > z)$$
where $p_L = \mathbb{P}(Z_i < z)$ and $p_R = \mathbb{P}(Z_i > z)$ denote the proportions of observations in the left and right child nodes, respectively. In a classical regression tree, the quantity of interest is the outcome $\xi_i = Y_i$, but in a causal tree, it is instead the treatment effect $\tau_{RD}$.

However, individual-level treatment effects $\tau_{RD}(X_i)$ cannot be directly observed, so causal forests instead rely on proxy estimates using average group-level differences. Intuitively, this can be understood as maintaining a separation between the treated and untreated samples in each node, enabling the tree structure to capture HTEs through local differences in group means. \citet{wager_estimation_2018} describe this as a form of data-driven population stratification, where each leaf ``acts as though [it] had come from a randomized experiment" restricted to individuals within that particular subgroup.

To ensure valid inference and prevent overfitting in the resulting tree, the key detail is a sample-splitting procedure known as honesty, where the data is randomly partitioned into two subsamples: one to construct the tree (i.e., determine splits), and another to then estimate treatment effects within the terminal leaves. The major theoretical results of \citet{athey_generalized_2019}, including consistency and asymptotic normality, rely on honesty along with several additional structural constraints, as specified below.

\begin{specification}[Athey et al., 2019]
All trees are symmetric, in that their output is invariant to permuting the indices of training examples; make balanced splits, in the sense that every split puts at least a fraction $\omega$ of the observations in the parent node into each child, for some $\omega > 0$; and are randomized in such a way that, at every split, the probability that the tree splits on the $j$-th feature is bounded from below by some $\pi > 0$. The forest is honest and built via subsampling with subsample size $s$ satisfying $s/n \rightarrow 0$ and $s \rightarrow \infty$.
\end{specification}

Together, these conditions ensure the resulting forest is an ensemble of weak learners that are both sufficiently diverse (due to randomization and subsampling) and statistically stable (due to honesty and balanced splits), allowing for valid pointwise confidence intervals.

To improve the empirical performance of causal forests, \citet{athey_generalized_2019} incorporate local centering through the double/debiased machine learning (DML) framework of \citet{chernozhukov_doubledebiased_2018}. This approach yields smoother and more robust estimates by separately regressing out the influence of the baseline covariates on both the treatment and the outcome. In practice, this is implemented by first fitting regression forests on $Y \sim X$ and $W \sim X$ to estimate the baseline risk $\widehat{Y} = \mathbb{E}[Y|X]$ and propensity score $\widehat{W} = \mathbb{E}[W|X]$. The main causal forest algorithm is then applied to the residualized outcome $\widetilde{Y} = Y - \widehat{Y}$ and residualized treatment $\widetilde{W} = W - \widehat{W}$.

%An important theoretical insight underlying DML is the concept of Neyman orthogonality, which ensures that the final treatment effect estimate is insensitive to small misspecifications in the initial regression models. In the context of causal forests, the baseline risk and propensity score forests are not of direct inferential interest, so they are considered nuisance functions, collectively denoted by $\eta = \{\widehat{Y}, \widehat{W}\}$. For a moment function $\psi(W, Y, X; \tau, \eta)$ aimed at estimating the treatment effect $\tau$, Neyman orthogonality requires that
%$$\left. \frac{\partial}{\partial \eta} \mathbb{E}[\psi(W, Y, X; \tau, \eta)] \right|_{\eta = \eta_0} = 0,$$
%meaning that small perturbations in $\eta$ around the true value $\eta_0$ do not affect the first-order expectation of the score function. In other words, the estimator for $\tau$ is doubly robust: it remains consistent even if one of the nuisance functions is misspecified, provided the other is accurately estimated. This property underpins the stability of orthogonalized causal forest estimates in high-dimensional or flexible settings.

In recent years, causal forests have gained widespread adoption across the biomedical and social sciences \citep{davis_using_2017, athey_estimating_2019, raghavan_generalizability_2022}. The current state-of-the-art implementation is the generalized random forests (\texttt{grf}) package in \texttt{R} \citep{tibshirani_grf_2023}, which extends forest-based inference to a broad class of statistical problems, including HTE estimation. 

\subsection{Model-based forests}

Meanwhile, model-based forests \citep{zeileis_model-based_2008, seibold_model-based_2016} explicitly specify a parametric model with the outcome as a function of the treatment and covariates
$$Y_i = \mu(X_i) + W_i \tau(X_i) + \epsilon_i.$$
In this setting, $\mu(X)$ represents the prognostic effect of baseline covariates that directly impact the outcome, while $\tau(X)$ is the predictive effect of covariates that influence treatment efficacy. A given covariate can be both prognostic and predictive.

For this model, define an objective function $\Psi((Y, X), \theta)$, such as the negative log-likelihood, with respect to parameters $\theta = (\mu, \tau)$. The model is fitted by minimizing this,  
$$\arg\min_{\theta} \sum_{i=1}^{n} \Psi((y_i, x_i), \theta),$$  
or equivalently by solving the score equation
$$\sum_{i=1}^{n} \frac{\partial \Psi((y_i, x_i), \theta)}{\partial \theta} = \sum_{i=1}^{n} \psi((y_i, x_i), \theta) = 0$$  
where $\psi$ is the score function, the gradient of the objective function with respect to the parameters \citep{seibold_model-based_2016}.

The key insight of model-based forests is that covariates associated with heterogeneity induce instabilities in the parameter estimates, and that this can be measured directly via score functions, which quantify how much each individual's data influences the estimation of $\mu$ and $\tau$. Specifically, define the partial score functions as
$$\psi_{\mu}((y, x), \theta) = \frac{\partial \Psi((y, x), \theta)}{\partial \mu} \;\; \text{and} \;\; \psi_{\tau}((y, x), \theta) = \frac{\partial \Psi((y, x), \theta)}{\partial \tau}.$$ 
If the true baseline effect $\mu$ is constant across individuals, then $\psi_{\mu}$ should be independent of any covariates; similarly, if the true treatment effect $\tau$ is homogeneous, then $\psi_{\tau}$ should be independent of all partitioning variables.

To detect covariates associated with heterogeneity, two hypothesis tests of independence can therefore be conducted for each candidate split variable $Z_j$ (where the splitting variables $Z$ and model variables $X$ are often the same, but are not required to be)
\begin{align*}
    H_{\mu, j}^{0}: \psi_{\mu}((y, x), \hat{\theta}) \perp Z_j, \quad j = 1, \dots, J\\
    H_{\tau, j}^{0}: \psi_{\tau}((y, x), \hat{\theta}) \perp Z_j, \quad j = 1, \dots, J.
\end{align*}
Out of these $2 \times J$ tests, the partitioning variable with the smallest p-value (subject to a certain threshold) is selected, indicating the strongest dependence between the covariate and changes in the prognostic or predictive effect.

Model-based forests build upon older approaches, such as RECPAM \citep{ciampi_recpam_1988} and GUIDE \citep{loh_regression_2002}, which conduct model comparison within each node of a tree to identify the optimal splitting variable. However, these approaches were typically applied to the general regression or classification setting, rather than to causal HTE estimation. In recent years, model-based forests have been extended to several other important applications, including individualized modeling \citep{seibold_individual_2018} and observational data integration \citep{dandl_heterogeneous_2024}. 

\section{Relative risk causal forests}

Forest-based methods are fundamentally characterized by the rule that determines whether and where to split the data at each node. In standard causal forests, this rule targets heterogeneity in the absolute risk because it aims to maximize variation in $\tau_{RD}(X_i)$. To instead target heterogeneity in the relative risk, we modify this splitting criterion to focus on variation in $\tau_{RR}(X_i)$. We begin by proposing a general alternative node-splitting procedure based on exhaustive generalized linear model (GLM) comparison, and then specialize it to the relative risk setting. Our implementation modifies the open-source \texttt{grf} package in \texttt{R}, and is available at \texttt{https://github.com/vshirvaikar/rrcf}.

\subsection{Forest construction}

Recall that at any given parent node, we observe a set of outcomes $Y_i$, covariates $X_i$, and binary treatment indicators $W_i \in \{0, 1\}$. To evaluate a candidate split on covariate $Z$ at threshold $z$, we first define a binary split indicator $S_i = \mathds{1}\{Z_i > z\}$ which denotes whether observation $i$ would fall to the left or right of the proposed split. We then fit the GLM
\begin{equation} \label{eq:glmsplit}
Y_i \sim X_i + W_i + S_i + W_i \cdot S_i
\end{equation}
Here, the $X_i$ terms adjust for baseline covariate effects, $W_i$ controls for the average treatment effect throughout the parent node, and $S_i$ controls for the main effect of the candidate split. The interaction term $W_i \cdot S_i$ then captures the difference in treatment effects between the two sides of the split -- in other words, the degree of treatment effect heterogeneity induced by the partition. We repeat this procedure for each candidate split, and select the one with the smallest $p$-value (or most extreme test statistic) for the interaction coefficient. While this approach is computationally intensive, it is easily parallelizable, and designed specifically for long-term clinical trial analysis where runtime is not a primary concern.

The motivation for GLM-based splitting is that we can target different treatment effect estimands by toggling the assumed response distribution and link function. For example, if we use linear regression (Gaussian distribution with an identity link), the estimated effect corresponds to an absolute risk difference. If we use logistic regression (binomial distribution with a logit link), the estimated effect represents a log-odds ratio. To target the relative risk ratio, we adopt Poisson regression with a log link, which models the response as
$$Y_i \sim \text{Poisson}(\lambda_i), \text{ with } \log(\lambda_i) = \beta_0 + \sum_{k} \beta_k x_{ik}.$$
Although the Poisson distribution is typically used for count data, the model remains valid for binary outcomes when the focus is on estimating relative risks, because the log link directly models multiplicative effects on the outcome scale. In this setting, $\lambda_i$ can be interpreted as an intensity parameter, analogous to a hazard function in survival analysis, capturing the instantaneous rate at which the binary outcome transitions from 0 to 1.

%An alternative approach would be log-binomial regression, which assumes a binomial likelihood and also uses a log link. This approach is theoretically more appropriate for binary data, as it ensures that fitted probabilities remain within the unit interval. However, log-binomial models are known to suffer from convergence issues, especially in small samples or near the boundaries. Poisson regression, by contrast, enjoys stable estimation due to its canonical link function, and has been shown to perform similarly to the log-binomial approach in comparative studies \citep{petersen_comparison_2008, chen_comparing_2018}. Consequently, this approach is implemented in the relative risk causal forest.

We incorporate additional adjustments to mirror the orthogonalization strategy used in standard causal forests. In classical DML, both the outcome and treatment variables are residualized to remove variation explained by baseline covariates \citep{chernozhukov_doubledebiased_2018}. However, the outcome variable in our setting must remain non-negative and integer-valued for Poisson regression, so we cannot use $\widetilde{Y} = Y - \widehat{Y}$ in Equation \ref{eq:glmsplit}. Instead, we still estimate a baseline risk model of the form $Y \sim X$, but then use the linear predictor 
$$\hat{\nu} = \log(\widehat{Y}) = \log(\mathbb{E}[Y \mid X])$$ 
to replace $X$ in Equation \ref{eq:glmsplit}, yielding a univariate adjustment for baseline outcome risk under the log link. While the functional form of this model is flexible, we use Poisson regression and extract the fitted linear predictor as $\hat{\nu} = X\hat{\beta}$. In the resulting GLM, $\hat{\nu}$ typically has a coefficient close to one, reflecting its role in capturing the background log-risk. This substitution also provides a computational benefit: it reduces the dimensionality of the repeated GLM fits from $p+3$ covariates to exactly four.

In the case of randomized controlled trial (RCT) data, where treatment is unconfounded by design, the adjusted model becomes
$$Y_i \sim \hat{\nu}_i + W_i + S_i + W_i \cdot S_i.$$
In observational or mixed designs, however, additional adjustment is required to mitigate confounding. Following the application of DML used in causal forests, we estimate the propensity score via a regression forest on $W \sim X$, yielding $\widehat{W} = \mathbb{E}[W \mid X]$, and replace the treatment indicator with the residualized treatment $\widetilde{W} = W - \widehat{W}$. The final model used for evaluating candidate splits in such settings is then
$$Y_i \sim \hat{\nu}_i + \widetilde{W}_i + S_i + \widetilde{W}_i \cdot S_i.$$

We implement our modified splitting criterion within the \texttt{grf} package in \texttt{R}, which relies on a \texttt{C++} backend for computational efficiency and scalability via the \texttt{Rcpp} framework \citep{eddelbuettel_rcpp_2011}. The GLM fits are performed in \texttt{C++} using iteratively reweighted least squares (IRLS), a standard algorithm for maximum likelihood estimation in GLMs. 
%IRLS solves a sequence of weighted least squares problems, where both the weights and the working response are updated based on the current parameter estimates. In our implementation, we use Householder QR decomposition at each step for improved numerical stability. Convergence is assessed via the $L_2$ norm of successive coefficient updates, and we terminate early if convergence stalls or fails. Upon convergence, we extract the final coefficient vector and its estimated covariance matrix to compute standard errors, and select the candidate split with the largest absolute $t$-statistic for the treatment-split interaction term. 
If no candidate split yields a converged model, we stop splitting. This commonly occurs in small subsamples or when outcomes are highly imbalanced, mirroring standard stopping behavior in classical random forests.

Because our method is implemented within the existing \texttt{grf} framework, it inherits the key structural properties described in Specification 1, including honesty and subsampling. We also retain the randomized selection of candidate splitting variables at each node, where the number of features considered is drawn from $\min\{\max\{\text{Poisson}(m), 1\}, k\}$ with tuning parameter $m > 0$, ensuring that every feature has a strictly positive probability of being selected \citep{denil_narrowing_2014}. The GLM-based splitting rule is symmetric (invariant to the ordering of training observations) and balanced (ensuring that each split places a nonzero fraction of samples from the parent node into each child). As a result, the theoretical guarantees established by \citet{athey_estimating_2019} under Specification 1, including consistency and asymptotic normality, continue to apply to our method.

\subsection{Treatment effect estimation}

With a trained forest in hand, we now consider how to estimate the treatment effect at a new test point $x$. Classical random forests frame prediction as an ensemble procedure: for each tree, the test point is passed down to a terminal leaf, the average outcome of training points in that leaf is computed, and the leaf-level means are aggregated across trees. 
%This yields the estimate
%$$\hat{\mu}(x) = \frac{1}{B} \sum_{b=1}^B \left( \frac{\sum_{X_i \in L_b(x)} Y_i}{\#\{X_i \in L_b(x)\}} \right),$$
%where $L_b(x)$ denotes the set of training samples that fall into the same leaf as $x$ in tree $b$, and $B$ is the total number of trees.

However, computational implementations of causal forests, including the \texttt{grf} package, recast this aggregation step as an adaptive nearest-neighbor procedure \citep{tibshirani_grf_2023}. Rather than computing tree-level predictions, these methods directly assign a weight to each training point based on how frequently it co-occurs in a leaf with $x$ across the forest. Let the resulting forest weights be denoted by $\alpha_i(x)$, defined as
$$\alpha_i(x) = \frac{1}{B} \sum_{b=1}^B \frac{\mathds{1}(X_i \in L_b(x))}{\#\{X_i \in L_b(x)\}},$$
where $L_b(x)$ denotes the set of training samples that fall into the same leaf as $x$ in tree $b$, $B$ is the total number of trees, and $\sum_{i=1}^n \alpha_i(x) = 1$. This formulation defines a kernel centered at $x$, allowing the forest to be interpreted as a locally weighted estimator over training examples. For causal applications, this enables smoother and more stable estimates of treatment effects, particularly when leaf sizes or class proportions vary across trees.

%$$ \hat{\tau}_{RD}(x) = \frac{\sum_{X_i \in L(x), W_i = 1} Y_i}{\#|X_i \in L(x), W_i = 1|} - \frac{\sum_{X_i \in L(x), W_i = 0} Y_i}{\#|X_i \in L(x), W_i = 0|}$$
%$$ \hat{\tau}_{RR}(x) = \left. \frac{\sum_{X_i \in L(x), W_i = 1} Y_i}{\#|X_i \in L(x), W_i = 1|} \middle/ \frac{\sum_{X_i \in L(x), W_i = 0} Y_i}{\#|X_i \in L(x), W_i = 0|} \right. $$

The adaptive weighting setup enables efficient computation of final estimates using pre-computed forest-wide statistics. In causal forests, this takes the form of a two-stage least squares estimate, motivated by the equivalence between causal and instrumental forests where the treatment assignment vector serves as the instrument \citep{athey_generalized_2019}. However, this cannot be directly extended to the relative risk setting, as the estimand is multiplicative rather than additive. Instead, we use the forest weights $\alpha_i(x)$ to compute weighted averages of outcomes within each treatment group. Specifically, for each test point $x$, we estimate the conditional risk under treatment and control as
$$\hat{\mu}_1(x) = \frac{\sum_{W_i = 1} \alpha_i(x) Y_i}{\sum_{W_i = 1} \alpha_i(x)}, \quad
\hat{\mu}_0(x) = \frac{\sum_{W_i = 0} \alpha_i(x) Y_i}{\sum_{W_i = 0} \alpha_i(x)}$$
The absolute risk estimate returned by a standard causal forest is then
$\hat{\tau}_{RD}(x) = \hat{\mu}_1(x) - \hat{\mu}_0(x)$
and so we can instead define the relative risk estimate as
$\hat{\tau}_{RR}(x) = \hat{\mu}_1(x) \big/ \hat{\mu}_0(x).$

\subsection{Omnibus testing}

Our final step is to evaluate whether the forest's individual-level relative treatment effect predictions $\hat{\tau}_{RR}(X_i)$ are well-calibrated and statistically significant. To this end, we develop an omnibus test for the detection of overall heterogeneity, extending the calibration test proposed by \citet{tibshirani_grf_2023} to the multiplicative setting.

The \texttt{grf} package assesses the overall quality of a forest via a simple linear fit on held-out data. Let $\widetilde{Y}$ and $\widetilde{W}$ denote the residualized outcome and treatment vectors as before, and let $\hat{\tau}_{RD}(X_i)$ represent the individual predicted absolute treatment effects. Define $\bar{\tau}_{RD}(X) = \mathbb{E}[\hat{\tau}_{RD}(X_i)]$ as the mean prediction across the test set. The test fits the linear model
\begin{equation} \label{eq:grfomni}
    \widetilde{Y}_i \sim \alpha\, \widetilde{W}_i\, \bar{\tau}_{RD}(X) + \beta\, \widetilde{W}_i\, \left(\hat{\tau}_{RD}(X_i) - \bar{\tau}_{RD}(X)\right),
\end{equation}
where $\alpha$ indicates whether the mean forest prediction is centered and $\beta$ indicates whether the individual HTE estimates are well-calibrated. A ``correct'' forest would have $\alpha = \beta = 1$, and Equation \ref{eq:grfomni} would reduce to $\widetilde{Y}_i = \widetilde{W}_i \hat{\tau}_{RD}(X_i)$. The p-value on $\beta$ is then used as an omnibus test for absolute heterogeneity: a small p-value indicates that the individual predictions contain significant explanatory signal beyond the global average.

In the relative risk setting, we analogously predict $\hat{\tau}_{RR}(X_i)$ for each individual in a held-out test set. We begin with the baseline model $Y \sim X + W$, and evaluate the added contribution of an interaction term $W \log(\hat{\tau}_{RR}(X))$. This yields the model
\begin{equation} \label{eq:omnibus}
Y_i \sim X_i + W_i + W_i \log(\hat{\tau}_{RR}(X_i)),
\end{equation}
where the log link means that the final term becomes exactly $\hat{\tau}_{RR}(X_i)$ for the treatment group, and zero for the control group. The p-value on the final term therefore serves as an omnibus test for relative heterogeneity in the relative treatment effect: it measures whether the individual-level predictions provide significant improvement over a model with only main effects for the covariates and treatment. 

\section{Simulations}

We conduct a simulation study to assess the effectiveness of the relative risk causal forest in detecting HTEs, and to compare its statistical power against the \texttt{grf} absolute risk baseline. Full details on the data-generating mechanism are provided in Appendix A.

Data are generated using the frugal parameterization of \citet{evans_parameterizing_2024}, which enables flexible specification of a known HTE structure. We simulate seven covariates in total: three that are prognostic for the outcome but do not influence the treatment effect, and four that serve as predictive effect modifiers. A heterogeneity parameter $\rho$ controls the strength of effect modification, allowing us to start with no heterogeneity ($\rho = 0$) and progressively increase the signal. We consider both an RCT setting, where treatment is assigned independently of covariates, and an observational setting, where treatment depends on a subset of the covariates. Potential outcomes are generated as functions of all covariates under each treatment condition.

We compare the methods across 100 random seeds, for a range of sample sizes and heterogeneity levels, with forests comprised of 500 trees. We first report statistical power, defined as the proportion of simulations in which the omnibus test yields a p-value below 0.05, with each forest trained on 80\% of the given sample size and tested on the other 20\%. We also report the average variable importance assigned to the true predictive covariates ($C_1$ to $C_4$) in the forest, as a measure of whether each method successfully identifies the relevant sources of heterogeneity. Variable importance is computed using the built-in function in the \texttt{grf} package, which weights variables based on how frequently they are used for splits, adjusted for their depth in the tree. Results are provided in Appendix A, with complete metrics in Table \ref{tab:sims}.

In the RCT setting, Figure \ref{fig:1A} displays the power of each method, while Figure \ref{fig:1B} shows the average importance assigned to the true predictive variables. As expected, performance improves with increasing sample size and heterogeneity level. Across both metrics, the relative risk approach consistently outperforms the baseline causal forest. In non-null simulations, the relative risk causal forest achieves an average increase in power of 5.2\%, and assigns 5.5\% more importance to the true effect modifiers.

In the observational setting, Figure \ref{fig:2A} displays power, while Figure \ref{fig:2B} compares key variable importance. The relative risk approach again uniformly outperforms the baseline causal forest in both metrics. The difference in performance is larger than in the RCT experiment, with an average power increase of 10.3\% in non-null simulations, and 9.5\% more importance assigned to the true effect modifiers. These results suggest that the relative risk forest is able to effectively adjust for differences in treatment propensity, making it a promising candidate for observational data applications.

\section{Data analysis}

In this section, we apply relative risk causal forests to the analysis of real-world data from the Liraglutide Effect and Action in Diabetes: Evaluation of Cardiovascular Outcome Results (LEADER) clinical trial. LEADER was initiated in 2010 to evaluate the benefit of liraglutide, a glucagon-like peptide 1 analogue, in the treatment of patients with type 2 diabetes \citep{marso_liraglutide_2016}. 9,340 patients underwent randomization with a median follow-up time of 3.8 years. The primary outcome (MACE, or major adverse cardiovascular events) occurred in significantly fewer patients in the treatment group --- 13.0\% compared to 14.9\%, with a hazard ratio of 0.87 (95\% confidence interval from 0.78 to 0.97).

From the LEADER dataset, 70 covariates were identified as potentially relevant, including a mix of demographic fields, vital signs, lab measurements, medical history flags, and medication flags. Details can be found in Appendix B. Only patients with no missing data across all covariates were used, resulting in a final analysis set of 8,750 observations, with a very similar outcome distribution to the complete data.

Including the primary MACE endpoint, 30 total primary and secondary outcomes were recorded. Recall that a key motivation for HTE estimation based on relative risk is its potential to improve power when baseline risk varies substantially across individuals. To operationalize this idea, we conduct a screening procedure across all available outcomes $Y^{(1)}, ..., Y^{(30)}$. We fit baseline logistic regression models $Y^{(j)} \sim X$ for $j = 1, ..., 30$ using the observed covariates, compute individual risk estimates $\hat{Y}^{(j)}_i$, and calculate the empirical variance of the predicted risk $\text{Var}(\hat{Y}^{(j)}_i)$ across the population. Table~\ref{tab:LEADER1} lists the ten outcomes with the highest baseline risk variance. The top five outcomes were selected for further analysis, with the primary MACE outcome ranking fifth. For reference, the baseline risk of MACE across the complete population ranges from 1.2\% to 62.4\% with a mean of 13.9\%. 

\begin{table}[]
\centering
\begin{tabular}{ccc}
\textbf{Outcome} & \textbf{Event Description} & \textbf{Baseline Risk Variance} \\ 
\hline
PRMACETM & Expanded MACE* & 14.4\% \\
HFDTHEVT & Composite heart failure or death & 12.9\% \\
MICROTM  & Microvascular event & 9.3\% \\
NEPHROTM & Secondary nephropathy event & 7.8\% \\
MACEEVTM & MACE* & 7.7\% \\
\hline
FRMASATM & MACE prior to 15th visit* & 7.6\% \\
ALDTHTM  & All-cause death & 7.2\% \\
PPEVENT  & MACE without pause over 120 days* & 5.2\% \\
OTR30EVT & MACE within 30 days of completion* & 4.9\% \\
EXMCHFTM & Heart failure requiring hospitalization* & 4.5\% \\
\hline
\end{tabular}
\caption{Baseline risk variances $\text{Var}(\hat{Y}_i)$ for the top ten outcomes from LEADER, using predicted risk from a logistic regression model. MACE refers to major adverse cardiovascular events; outcomes marked with an asterisk (*) require confirmation from an event adjudication committee.}
\label{tab:LEADER1}
\end{table}

For each of the top five outcomes, we perform five-fold cross-validation using forests composed of 2,000 trees. In each fold, 80\% of the data are used to train the forest, and relative risk coefficients $\hat{\tau}_{RR}$ are predicted for the remaining 20\%. The out-of-fold predictions are concatenated to yield a vector of estimates across the full dataset of 8,750 observations, and the omnibus test from Equation \ref{eq:omnibus} is applied. Table \ref{tab:LEADER2} summarizes the results: for the \texttt{grf} absolute risk forest, none of the five $p$-values indicate useful findings, but for the relative risk approach, two are suggestive at $\alpha = 0.1$, and one outcome is significant at $\alpha = 0.05$. We focus on this outcome (HFDTHEVT, indicating composite heart failure or death) for subsequent analysis.

\begin{table}[]
\centering
\begin{tabular}{cccc}
\multirow{2}{*}{\textbf{Outcome}} & \multirow{2}{*}{\textbf{Event Description}} & \multicolumn{2}{c}{\hspace{-0.7em}\textbf{Omnibus p-value}} \\
& & \multicolumn{1}{c}{\textbf{GRF}} & \multicolumn{1}{c}{\textbf{RRCF}} \\ 
\hline
PRMACETM & Expanded MACE* & 0.724 & 0.082 \\
HFDTHEVT & Composite heart failure or death & 0.921 & \textbf{0.027} \\
MICROTM  & Microvascular event & 0.904 & 0.512   \\
NEPHROTM & Secondary nephropathy event & 0.238 & 0.072 \\
MACEEVTM & MACE* & 0.163 & 0.908 \\
\hline
\end{tabular}
\caption{Cross-validated omnibus test results for the top five outcomes from LEADER, comparing p-values for absolute and relative risk causal forests. MACE refers to major adverse cardiovascular events; outcomes marked with an asterisk (*) require confirmation from an event adjudication committee. The bolded p-value (composite heart failure or death, for the relative risk causal forest) indicates the only significant finding at $\alpha=0.05$.}
\label{tab:LEADER2}
\end{table}

To translate our HTE findings into clinically relevant insights, a key question is which covariates or interactions drive variation in treatment effectiveness, as captured by the fitted relative risk estimates $\hat{\tau}_{RR}(X_i)$ from the causal forest. For the composite heart failure or death outcome, we focus on two subgroups: (1) the decile of patients with the greatest predicted benefit (those with $\hat{\tau}_{RR}(X_i) < 0.815$), which we refer to as the \emph{high-benefit group}; and (2) the subset of patients for whom treatment is predicted to be harmful (those with $\hat{\tau}_{RR}(X_i) > 1$), comprising 7.6\% of the population, which we refer to as the \emph{low-benefit group}. The remaining 82.4\% of individuals are designated as the reference population.

To characterize each subgroup, we perform two-sample $t$-tests comparing each covariate to the reference population. Table \ref{tab:LEADER3} lists the five covariates with the most significant differences (indicated by $p$-value) between each subgroup and the reference. Overall, the high-benefit group appears less healthy, with elevated blood pressure and cholesterol, while the low-benefit group is comparatively healthier. 

\begin{figure}
\centerline{\includegraphics[width=\textwidth]{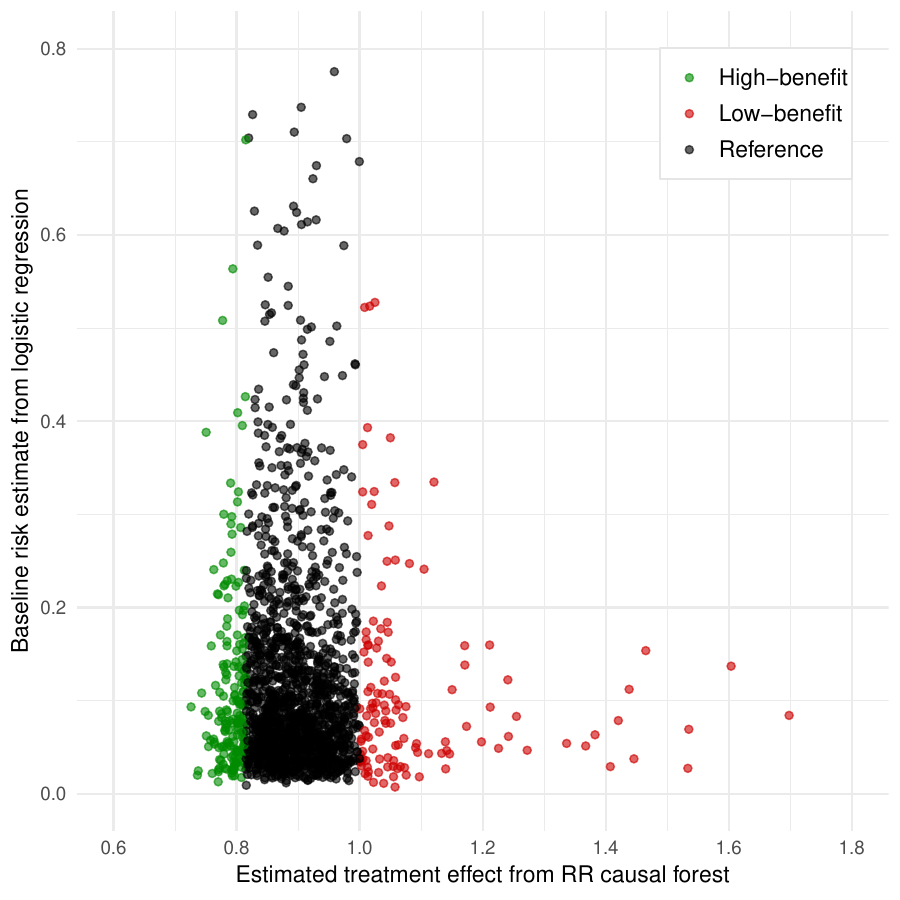}}
\caption{Baseline risk estimates $\hat{Y}_i$ from logistic regression against relative treatment effects $\hat{\tau}_{RR}(X_i)$ from the causal forest. Points are color-coded by subgroup.\label{fig:3}}
\end{figure}

However, the relationship is not explained by overall health alone. Figure \ref{fig:3} plots baseline risk estimates $\hat{Y}_i$ from the initial logistic regression against relative treatment effects $\hat{\tau}_{RR}(X_i)$ from the relative risk causal forest. There is a slight negative trend, but the correlation is weak ($\rho = -0.053$), suggesting that more complex interactions underlie heterogeneity.

\begin{table}[]
\centering
\begin{tabular}{cccc}
& \multicolumn{2}{c}{\textbf{Subgroup Mean}} & \\
\textbf{Covariate} & \textbf{Reference} & \textbf{High-Benefit} & \textbf{Percent Difference} \\ 
\hline
Systolic blood pressure & 135 & 147 & +8.35\% \\
Total cholesterol & 4.37 & 5.13 & +17.3\% \\
LDL cholesterol & 2.30 & 2.90 & +26.1\% \\
Sex (Male=1) & 0.66 & 0.39 & -40.9\% \\
Hemoglobin & 8.53 & 8.02 & -6.02\% \\
\hline
& \textbf{Reference} &\textbf{Low-Benefit} & \\ 
\hline
Total cholesterol & 4.37 & 3.76 & -13.9\% \\
Platelets & 269 & 222 & -17.6\% \\
LDL cholesterol & 2.30 & 1.89 & -17.8\% \\
Sex (Male=1) & 0.66 & 0.85 & +28.5\% \\
Systolic blood pressure & 135 & 128 & -5.61\% \\
\hline
\end{tabular}
\caption{Covariates with the five most significant differences, measured by $p$-value, from the reference population to the high-benefit (top) and low-benefit (bottom) subgroups. All $p$-values are significant at $\alpha = 10^{-25}$.}
\label{tab:LEADER3}
\end{table}

In addition, simulation results suggest that relative risk causal forests more effectively split on predictive covariates which drive treatment effect heterogeneity. To explore this, we compare variable importance between the absolute and relative risk approaches. Table~\ref{tab:LEADER4} displays the five covariates with the largest gains and losses in importance under the relative risk model, averaged across five training folds.

White blood cell and platelet counts show the greatest increases in importance, suggesting that blood-related biomarkers, particularly those linked to immune or inflammatory activity, play a central role in driving heterogeneity on the relative scale. However, these effects may be underrepresented in absolute risk models, possibly because their influence is stronger among individuals with lower baseline risk. Conversely, two of the five largest decreases in importance involve estimated glomerular filtration rate (eGFR), a marker of kidney function. While \citet{marso_liraglutide_2016} identified eGFR as a potential effect modifier, our findings suggest that its role may be overstated when heterogeneity is analyzed in absolute terms, and less relevant under a relative risk framework.

\begin{table}[]
\centering
\begin{tabular}{cccc}
\multirow{2}{*}{\textbf{Covariate}} & \multicolumn{2}{c}{\textbf{Importance}} & \multirow{2}{*}{\textbf{Difference}} \\
& \multicolumn{1}{c}{\textbf{GRF}} & \multicolumn{1}{c}{\textbf{RRCF}} & \\ 
\hline
White blood cells & 3.39\% & 14.58\% & +11.19\% \\
Platelets & 3.17\% & 8.01\% & +4.84\% \\
Amylase & 2.27\% & 3.60\% & +1.33\% \\
Potassium & 2.16\% & 2.86\% & +0.70\% \\
Sodium & 0.87\% & 1.55\% & +0.68\% \\
\hline
eGFR by CKD-EPI & 4.52\% & 1.87\% & -2.65\% \\
Hematocrit & 4.51\% & 2.52\% & -1.99\% \\
Systolic blood pressure & 5.62\% & 3.79\% & -1.83\% \\
eGFR by MDRD & 3.83\% & 2.15\% & -1.68\% \\
Age & 4.47\% & 2.81\% & -1.66\% \\
\hline
\end{tabular}
\caption{Covariates with the five largest gains (top) and five largest losses (bottom) in variable importance between absolute and relative risk forests, averaged across five training folds. eGFR refers to estimated glomerular filtration rate, a marker of kidney function, which is estimated according to two different formulas in the provided data.}
\label{tab:LEADER4}
\end{table}

\section{Conclusion}

In this study, we present an adaptation of causal forests that specifically targets heterogeneity in the relative Risk Ratio (RR). The RR is an important clinical measure for several reasons, including its ability to generalize across populations. HTE discovery methods that only target the absolute Risk Difference (RD) may overlook critical sources of heterogeneity on the RR scale, especially in settings with large variation in individual baseline risk. 

The modification is based on an alternative splitting rule for the data in a causal forest that uses exhaustive generalized linear model (GLM) comparison. Specifying Poisson regression as the GLM of choice allows the RR to be targeted as the quantity of interest. We validate the forests' performance on simulated data, and explore real-world data from the LEADER cardiovascular outcome trial. These results suggest the RR adjustment can improve the power of causal forests to identify heterogeneity.

\section*{Acknowledgements}
We thank James Watson and Kate Ross for their helpful revisions and feedback. We also thank George Nicholson, Fabian Falck, and collaborators at Novo Nordisk for technical advice and guidance. VS is supported by the EPSRC Centre for Doctoral Training in Modern Statistics and Statistical Machine Learning (EP/S023151/1) and Novo Nordisk. CH is supported by Novo Nordisk.

%\setlength{\bibitemsep}{0.5\baselineskip}
%\setlength{\bibhang}{0pt}
%\printbibliography[heading=none]

\bibliography{export-data}

\newpage
\appendix

\setcounter{figure}{0}
\setcounter{table}{0}
\renewcommand\thefigure{A\arabic{figure}}
\renewcommand\thetable{A\arabic{table}}

\section*{Appendix A: Simulation details and results}

For a meaningful simulation study with a known true treatment effect, it is critical to simulate data from a correctly specified data-generating process. While this is relatively straightforward in the case of additive treatment effects, greater care is required when modeling potential outcomes on a relative scale. Following \citet{lin_data_2024}, we adopt the frugal parameterization proposed by \citet{evans_parameterizing_2024} to define and simulate from marginal structural models. This framework allows for precise control over the HTE structure by separately specifying the key components of the joint distribution. Specifically, the joint distribution of covariates, treatment, and potential outcomes is decomposed into three distinct components: (1) the marginal causal quantity of interest, i.e., the HTE function; (2) the joint distribution of the treatment and covariates; and (iii) the dependence between the outcome and covariates conditional upon the treatment. 

We simulate seven covariates in total. $X_1$, $X_2$, and $X_3$ are prognostic for the outcome $Y$ but do not modify the treatment effect, while $C_1$, $C_2$, $C_3$, and $C_4$ are predictive effect modifiers. These variables are a mix of discrete and continuous as specified below:
$$X_1 \sim N(0,1) \quad \quad X_2 \sim \operatorname{Gamma}(0.1 + 0.2X_1, 1) \quad \quad X_3 \sim \operatorname{Beta}(0.1 + 0.1X_1, 1)$$
$$C_1 \sim \operatorname{Bernoulli}(0.5) \quad \quad C_2 \sim \operatorname{Bernoulli}(\operatorname{expit}(-2 + C_1))$$
$$C_3 \sim N(0.1C_1C_2,1) \quad \quad C_4 \sim t_{20}(0.1 C_1,0.1)$$

In the RCT setup, treatment is simulated as $W \sim \operatorname{Bernoulli}(0.5)$. In the observational setup, we simulate $W \sim \operatorname{Bernoulli}(\operatorname{expit}(-1 + 2X_1 + 2X_3 - C_2))$.

Potential outcomes are drawn from the marginal $Y(w) \sim \operatorname{Bernoulli}(\mu_y)$, where
\begin{equation*}
    \operatorname{log}(\mu_y) = -2 + 0.3C_1 + 0.4 \sin(C_4) + W \left\{-0.2 + \rho \left(C_1 + C_2 + \mathbb{I}(C_3 > 0) + C_4^2\right) \right \}
\end{equation*}
This results in a CATE given by
\begin{equation*}
\tau_{RR}(\boldsymbol{c}) = \exp \left \{-0.2 + \rho \left(C_1 + C_2 + \mathbb{I}(C_3 > 0) + C_4^2\right)\right \}
\end{equation*}
where $\rho$ controls the degree of heterogeneity. When $\rho = 0$, there is no heterogeneity and the treatment effect is homogeneous at a factor of $\exp(-0.2) \approx 0.819$. We vary $\rho$ across $\{0, 0.25, 0.5, 0.75\}$, resulting in a progressively stronger signal.

A pair-copula construction is used to flexibly encode the dependency between the non-modifier covariates and the outcome. Specifically, we employ a Gaussian copula to model the $X_1-Y$ dependency with correlation depending on $C_2$; a Clayton copula to capture higher-tail dependency for $X_2-Y$; and a Gumbel copula to introduce lower-tail dependency for $X_3-Y$. We simulate these complex distributions to better reflect real-world scenarios, where data often deviates from simple Gaussian and linear models.

\begin{figure}
\centerline{\includegraphics[width=\textwidth]{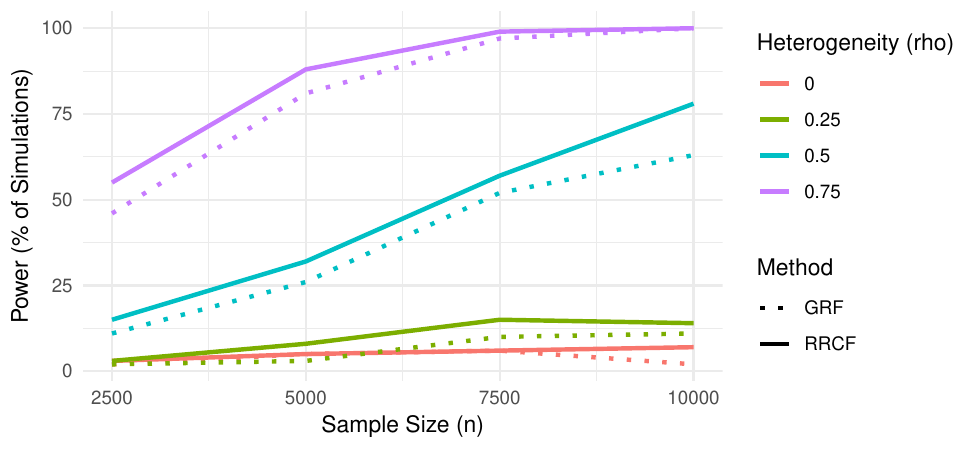}}
\caption{Power (proportion of trials where omnibus test p-value on additional $W_i \log(\hat{\tau}_{RR}(X_i))$ term was significant) across 100 RCT simulations, as a function of sample size $n$ and heterogeneity level $\rho$. \label{fig:1A}}
\vspace{1cm}
\centerline{\includegraphics[width=\textwidth]{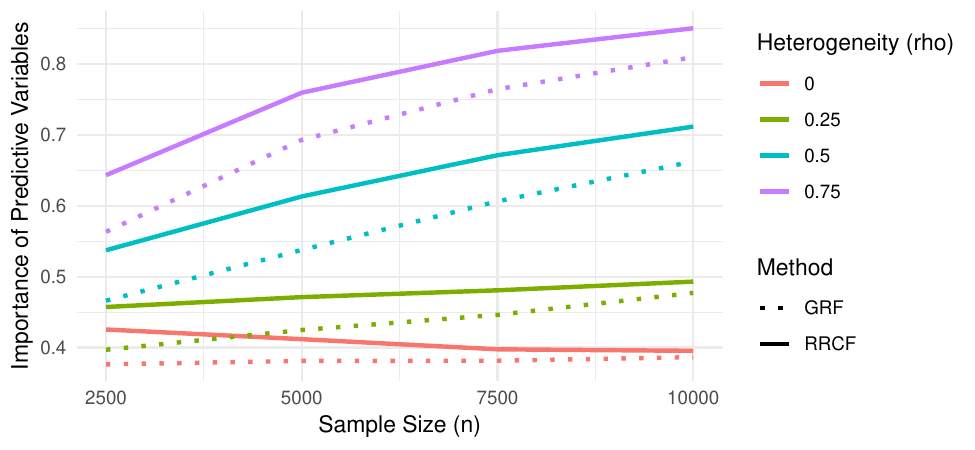}}
\caption{Average variable importance assigned to true predictive covariates ($C_1$ to $C_4$) across 100 RCT simulations, as a function of sample size $n$ and heterogeneity level $\rho$. \label{fig:1B}}
\end{figure}

\begin{figure}
\centerline{\includegraphics[width=\textwidth]{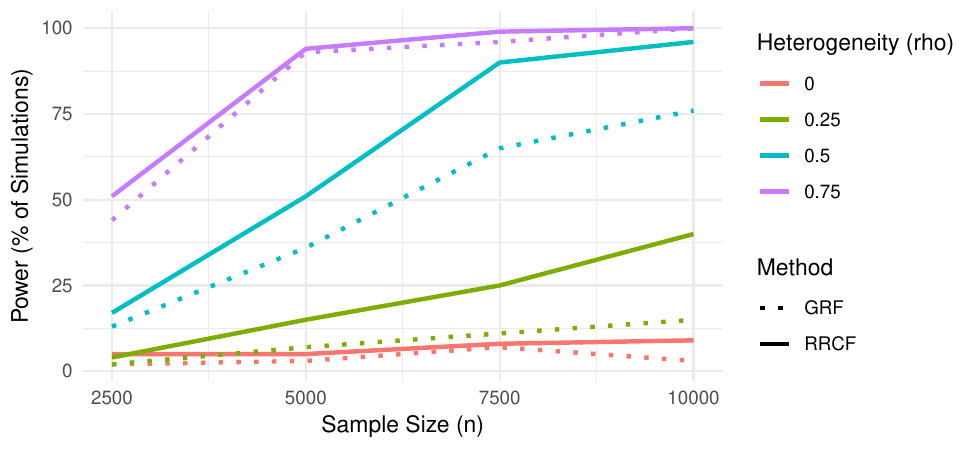}}
\caption{Power (proportion of trials where omnibus test p-value on additional $W_i \log(\hat{\tau}_{RR}(X_i))$ term was significant) across 100 observational simulations, as a function of sample size $n$ and heterogeneity level $\rho$. \label{fig:2A}}
\vspace{1cm}
\centerline{\includegraphics[width=\textwidth]{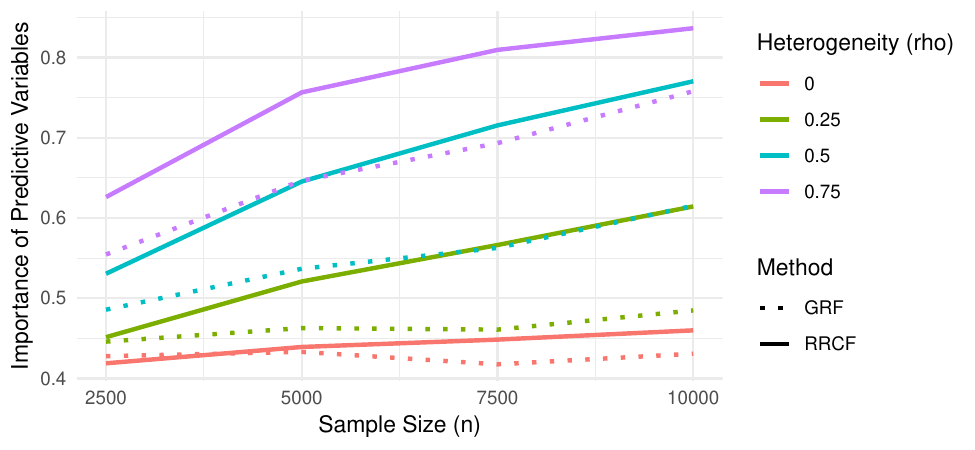}}
\caption{Average variable importance assigned to true predictive covariates ($C_1$ to $C_4$) across 100 RCT simulations, as a function of sample size $n$ and heterogeneity level $\rho$. \label{fig:2B}}
\end{figure}

\renewcommand{\arraystretch}{1.2}
\begin{table}[h]
\centering
\begin{tabular}{cccccccccc}
\hline
\textbf{} & \textbf{} & \multicolumn{4}{c}{\textbf{RCT data}} & \multicolumn{4}{c}{\textbf{Observational data}} \\ 
\hline
\textbf{} & \textbf{} & \multicolumn{2}{c}{\textbf{Power (\%)}} & \multicolumn{2}{c}{\textbf{Var. Imp. (\%)}} & \multicolumn{2}{c}{\textbf{Power (\%)}} & \multicolumn{2}{c}{\textbf{Var. Imp. (\%)}} \\ 
\hline
\textbf{HTE $\rho$}  & \textbf{$n$} & \textbf{GRF} & \textbf{RRCF} & \textbf{GRF} & \textbf{RRCF} & \textbf{GRF} & \textbf{RRCF} & \textbf{GRF} & \textbf{RRCF} \\ 
\hline
\multirow{4}{*}{\textbf{0}} & \textbf{2,500} & 2 & 3 & 37.6 & 42.5 & 2 & 5 & 42.8 & 41.9 \\
   & \textbf{5,000} & 5 & 5 & 38.1 & 41.2 & 3 & 5 & 43.3 & 43.9 \\
   & \textbf{7,500} & 6 & 6 & 38.1 & 39.7 & 7 & 8 & 41.8 & 44.8 \\
   & \textbf{10,000} & 2 & 7 & 38.6 & 39.5 & 3 & 9 & 43.1 & 46.0 \\
\hline
\multirow{4}{*}{\textbf{0.25}} & \textbf{2,500} & 2 & 3 & 39.7 & 45.7 & 2 & 4 & 44.6 & 45.1 \\
   & \textbf{5,000} & 3 & 8 & 42.5 & 47.1 & 7 & 15 & 46.3 & 52.1 \\
   & \textbf{7,500} & 10 & 15 & 44.6 & 48.1 & 11 & 25 & 46.1 & 56.6 \\
   & \textbf{10,000} & 11 & 14 & 47.7 & 49.3 & 15 & 40 & 48.5 & 61.4 \\
\hline
\multirow{4}{*}{\textbf{0.5}} & \textbf{2,500} & 11 & 15 & 46.6 & 53.7 & 13 & 17 & 48.6 & 53.0 \\
   & \textbf{5,000} & 26 & 32 & 53.8 & 61.3 & 36 & 51 & 53.7 & 64.5 \\
   & \textbf{7,500} & 52 & 57 & 60.6 & 67.1 & 65 & 90 & 56.3 & 71.5 \\
   & \textbf{10,000} & 63 & 78 & 66.3 & 71.2 & 76 & 96 & 61.5 & 77.0 \\
\hline
\multirow{4}{*}{\textbf{0.75}} & \textbf{2,500} & 46 & 55 & 56.3 & 64.3 & 44 & 51 & 55.4 & 62.6 \\
   & \textbf{5,000} & 81 & 88 & 69.3 & 76.0 & 93 & 94 & 64.7 & 75.6 \\
   & \textbf{7,500} & 97 & 99 & 76.5 & 81.9 & 96 & 99 & 69.3 & 80.9 \\
   & \textbf{10,000} & 100 & 100 & 81.0 & 85.1 & 100 & 100 & 75.8 & 83.6 \\
\hline
\end{tabular}
\caption{Complete numerical results for simulation experiment, including power (proportion of tests with omnibus p-value below 0.05) and variable importance (average weight assigned to true predictive covariates). \label{tab:sims}}
\end{table}

\clearpage

\section*{Appendix B: List of LEADER variables}

\begin{itemize}
    \item \textbf{6 demographic fields:} Age, Diabetes Duration, Sex, Race, Smoker, Previous Smoker
    \item \textbf{11 baseline vital signs:} Waist Circumference, Body Mass Index, Pulse, Systolic Blood Pressure, Diastolic Blood Pressure, HbA1c, HDL Cholesterol, LDL Cholesterol, Total Cholesterol, Triglycerides, Serum Creatinine
    \item \textbf{14 lab measurements:} Alanine Aminotransferase, Amylase, Bilirubin, Calcium, Estimated Glomerular Filtration Rate (eGFR) by the Chronic Kidney Disease-Epidemiology Collaboration (CKD-EPI) Formula, Estimated Glomerular Filtration Rate (eGFR) by the Modification of Diet in Renal Disease (MDRD) Formula, Potassium, Triacylglycerol Lipase, Sodium, Hematocrit, Hemoglobin, Platelets, Erythrocytes, Leukocytes
    \item \textbf{19 medical history flags:} Antihypertensive Therapy, Myocardial Infarction, Stroke, Stroke Sensitivity Analysis, Revascularization, Carotid Stenosis on Angiography, Coronary Heart Disease, Ischaemic Heart Disease, Chronic Heart Failure, Chronic Kidney Failure, Microalbuminuria or Proteinuria, Hypertension and Left Ventricular Hypertrophy, Left Ventricular Systolic and Diastolic Dysfunction, Ankle/Brachial Index, Cardiovascular High Risk, Cardiovascular Medium Risk, Diabetic Retinopathy, Diabetic Nephropathy
    \item \textbf{20 concomitant medication flags:} Insulin, Metformin, Sulfonylureas, Alpha Glucosinade Inhibitors, Thiazolidinediones, Glinides, Vitamin K Antagonists, Platelet Inhibitors, Other Antihypertensives, Thiazides, Thiazide-like Diuretics, Loop Diuretics, Aldosterone Antagonists, Beta-blockers, Calcium Channel Blockers, Angiotensin-Converting Enzyme Inhibitors, Angiotensin Receptor Blockers, Statins, Other Lipid Lowering Drugs, Ezetimibe
\end{itemize}

\end{document}